\documentclass{aa}
\usepackage{comment}
\usepackage{amsmath}
\usepackage{rotating}
\usepackage{multirow}
\usepackage{multicol}
\usepackage{hyperref}
\usepackage{multirow}
\usepackage{array}
\usepackage{longtable}
\usepackage{tablefootnote}
\usepackage{ulem}
\hypersetup{
    colorlinks=true,            
    linkcolor=black,             
    citecolor=blue,              
    urlcolor=blue               
}
\usepackage{graphicx}
\usepackage{txfonts}
\usepackage{hyperref}
\usepackage{academicons}
\usepackage{xcolor}
\begin{document}

   \title{MAISTEP - a new grid-based machine learning tool for inferring stellar parameters\\ 
   I. Ages of giant-planet host stars}

   \author{J. Kamulali
          \inst{1,2,3},
          B. Nsamba\inst{2,1,3}\href{https://orcid.org/0000-0002-4647-2068}{\includegraphics[scale=0.07]{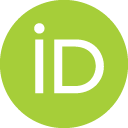}},
          V. Adibekyan\inst{3,4}\href{https://orcid.org/0000-0002-0601-6199}{\includegraphics[scale=0.07]{figures/ORCID_logo.png}},
          A. Weiss\inst{2}\href{https://orcid.org/0000-0002-3843-1653}{\includegraphics[scale=0.07]{figures/ORCID_logo.png}},
          T. L. Campante\inst{3,4}\href{https://orcid.org/0000-0002-4588-5389}{\includegraphics[scale=0.07]{figures/ORCID_logo.png}},
          \and
          N. C. Santos\inst{3,4}\href{https://orcid.org/0000-0003-4422-2919}{\includegraphics[scale=0.07]{figures/ORCID_logo.png}}
         }
          
   \institute{Department of Physics, Faculty of Science, Kyambogo University, P.O. Box 1, Kyambogo, Uganda\\
              \email{kamulali@mpa-garching.mpg.de}
         \and
             Max-Planck-Institut f\"{u}r Astrophysik, Karl-Schwarzschild-Str. 1, D-85748 Garching, Germany
         \and
            Instituto de Astrof\'{\i}sica e Ci\^{e}ncias do Espa\c{c}o, Universidade do Porto, Rua das Estrelas, 4150-762 Porto, Portugal
         \and
            Departamento de F\'{\i}sica e Astronomia, Faculdade de Ci\^{e}ncias da Universidade do Porto, Rua do Campo Alegre, s/n, 4169-007 Porto, Portugal
             }

   \date{\textbf{Received:}  03 December 2024  \textbf{Accepted:} 30 January 2025}
   \titlerunning{Ages of giant-planet host stars}
   \authorrunning{Kamulali et al.}
 
  \abstract
   {Our understanding of exoplanet demographics partly depends on their corresponding host star parameters. With the majority of exoplanet-host stars having only atmospheric constraints available, robust inference of their parameters (including ages) is susceptible to the approach used.   
   }
   {The goal of this work is to develop a grid-based machine learning tool capable of determining the stellar radius, mass, and age using only atmospheric constraints and to analyse the age distribution of stars hosting giant planets.}
   {Our machine learning approach involves combining four tree-based machine learning algorithms (Random Forest, Extra Trees, Extreme Gradient Boosting, and CatBoost) trained on a grid of stellar models to infer stellar radius, mass, and age using effective temperatures, metallicities, and $Gaia$-based luminosities. We perform a detailed statistical analysis to compare the inferences of our tool  with those based on seismic data from the APOKASC (based on global oscillation parameters) and LEGACY (based on individual oscillation frequencies) samples. Finally, we apply our tool to determine the ages of stars hosting giant planets.}
   { Comparing the stellar parameter inferences from our machine learning tool with those from the APOKASC and LEGACY, we find a bias (and a scatter) of  -0.5\% (5\%) and -0.2\% (2\%) in radius,  6\% (5\%) and -2\% (3\%) in mass, and  -9\% (16\%) and 7\% (23\%) in age, respectively. Therefore, our machine learning predictions are commensurate with seismic inferences.  When applying our model to a sample of stars hosting Jupiter-mass planets, we find the average age estimates for the hosts of Hot Jupiters, Warm Jupiters, and Cold Jupiters to be 1.98 Gyr, 2.98 Gyr, and 3.51 Gyr, respectively.}
   {Our machine learning tool is robust and efficient in estimating the stellar radius, mass, and age when only atmospheric constraints are available. 
   Furthermore, the inferred age distribution of giant-planet host stars confirms previous predictions -- based on stellar model ages for a relatively small number of hosts, as well as on the average age-velocity dispersion relation -- that stars hosting Hot Jupiters are statistically younger than those hosting Warm and Cold Jupiters.}
   
   \keywords{methods: statistical / stars: fundamental parameters / stars: solar-type / planets and satellites: general}

   \maketitle
%
\section{Introduction}

  Proper characterisation of exoplanet-host stars plays a vital role in our understanding of exoplanetary systems. This is based on the synergy between stellar and planetary science. For instance, the  planet's radius and mass, determined through transits and radial velocity analysis, rely on the known radius and mass of the host star, respectively \citep[e.g.,][]{seager2003unique,torres2008improved,mortier2013new,perryman2018exoplanet,hara2023statistical}. Stellar ages have also been used in investigating changes in the exoplanet demographics with time \citep[e.g.,][]{berger2020gaia,chen2023evolution}. They are also expected to be relevant in distinguishing between the different theories regarding the formation and evolution of Hot Jupiters \citep[][]{dawson2018origins}. 
  
 In order to determine robust stellar radius, mass, and age, various methods have been developed. \cite{torres2010accurate} devised polynomial functions based on a sample of non-interacting eclipsing binary systems with model-independent radii and masses available, and with measured parameters such as effective temperature, $T_{\text{eff}}$, surface gravity, log $g$, and metallicity, [Fe/H]. These relations have been adopted to uniformly estimate similar parameters of planet host stars \citep[e.g.,][]{santos2013sweet,hamer2019hot}. Further, \cite{moya2018empirical} used an extended sample of stellar targets with radius and mass estimates obtained through the analysis of detached eclipsing binary star systems, interferometric measurements, and seismic inferences. They reported 38 empirical relations for estimating mass and radius which were obtained from linear combinations of stellar density, $\rho$, luminosity, $L$, [Fe/H], $T_{\text{eff}}$, and log $g$. However, the relations by both \cite{torres2010accurate}  and \cite{moya2018empirical} cannot provide any information about the  stellar interior structure and ages.
 
 Stellar ages are typically inferred through empirical approaches, such as gyro-chronology \citep[e.g.,][]{angus2015calibrating,meibom2015spin, barnes2016rotation,claytor2020chemical}, chemical clocks \citep[e.g.,][]{da2012accurate,nissen2015high,nissen2016high,feltzing2016metallicity,titarenko2019ambre,mena2019abundance,espinoza2021consistency,moya2022stella}, and  stellar activity \citep[e.g.,][]{noyes1984rotation,mamajek2008improved}, as well as model-dependent techniques like isochrone fitting \citep[e.g., ][]{liu2000relative,sandage2003age,vandenberg2013ages,berger2020gaia} and asteroseismology \citep[e.g., ][]{silva2011constraining,silva2013stellar,campante2015ancient,campante2017weighing,silva2017standing,nsamba2018asteroseismic,nsamba2019nature,moedas2020asteroseismic,deal2021fundamental,moedas2024characterisation}. Each of these  methods works well in particular regions of the Hertzsprung-Russell (HR) diagram. For a comprehensive summary of these techniques, refer to the review by \cite{soderblom2010ages,soderblom2015stellar}. 
 
 The grid-based methods involve finding the best match of stellar observables to model observables through processes like iterative optimisation \citep{gai2011depth}, Markov Chain Monte Carlo \citep[MCMC;][]{bazot2012bayesian} sampling, and genetic optimisation algorithms \citep{metcalfe2009stellar}. In iterative optimisation, stellar models are specially constructed for the region of parameter space where the solution is believed to lie. A subsequent optimisation procedure is then carried out to find  the best-fitting model based on the available observable parameters of the star. MCMC is an augmentation of iterative technique allowing for an extensive  probabilistic  exploration of the parameter space and yields accurate estimates of stellar parameters with credible confidence intervals \citep[e.g.,][]{bazot2012bayesian,gruberbauer2012toward,lund2017asteroseismic,claytor2020chemical}. In a genetic optimisation algorithm, stellar tracks have to be run each time a new target is to be characterized. Additionally, the algorithm must be initialized from different starting points to ensure convergence  to a global  minimum. Thus, the effectiveness of these methods requires running dense grids of models.

 With large datasets, machine learning (ML) techniques have been efficiently utilised in the characterisation of stars. The application of ML algorithms for regression tasks in astrophysics dates back to the earlier works of \cite{pulone1997age}, in which an artificial neural network was built to determine stellar ages based on the position of the star on the HR diagram. Using a combination of classical and asteroseismic observations,  \cite{verma2016asteroseismic} and \cite{bellinger2016fundamental} characterised sun-like stars through the application of Neural networks and Random Forest, respectively. Despite the tremendous progress of ML codes in utilising both seismic and classical constraints to estimate stellar radius, mass, and age, the majority of stars do not have seismic data. This opens an opportunity to explore the effectiveness of ML codes in the absence of seismic data.
 
 \cite{moya2022stellar} used atmospheric constraints to determine the capabilities of various machine learning algorithms in inferring stellar mass and radius. They demonstrate that combining (stacking) the high-performing algorithms produces a more robust stellar model than any of the individual algorithms. In addition, they trained their ML algorithms on data from eclipsing binary stars, interferometry, and asteroseismology. However, only inferences of the stellar mass and radius were conducted in their work.

 In this study, we employ a stacking approach, similar to \cite{moya2022stellar}, but combine (stack) tree-based ensemble algorithms (RF, XT, CatBoost and XGBoost). Tree-based algorithms are computationally efficient (i.e., require relatively low computational resources) compared to other algorithms, such as neural networks \citep[][]{hastie2009elements}. In addition, they are effective in developing regression models that capture complex non-linear relationships \citep{hastie2009elements,baron2019machine}, particularly in linking observed stellar parameters to their fundamental properties (such as mass, radius, age, etc). To achieve this, we train our algorithms directly on grids of stellar models to determine not only the radius and the mass but also the stellar age using atmospheric constraints. Furthermore, this approach makes it possible to explore the impact of stellar model physics on the predictions made. The rest of the paper is structured as follows. In Sec.~\ref{maistep}, we present our ML tool and its underlying mechanisms. This is followed by an evaluation of the tool in Sec.~\ref{maistep_eval}. We then apply the technique to characterise planet host stars in Sec.~\ref{giant_planets}. Finally, we provide the summary and conclusions of our results in Sec.~\ref{summary_conclusion}.
  
\section{MAISTEP code}
\label{maistep}

\begin{figure*}[t]
        \centering
        \includegraphics[scale = 0.42]{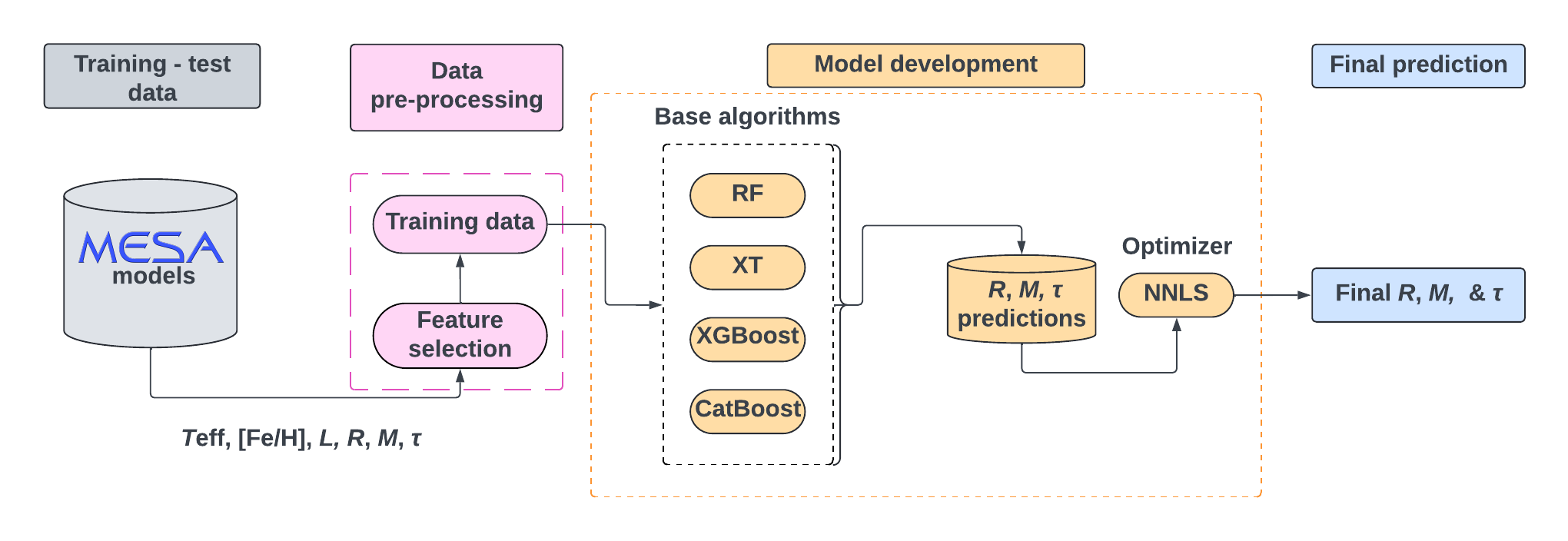}
            \caption{Schematic representation of MAISTEP. The training data undergoes pre-processing before being passed to the different base  algorithms under the model development phase, ultimately predicting the radius, mass, and age. See text for details.}
        \label{MAISTEP}
\end{figure*}

 Machine learning Algorithm for Inferring STEllar Parameters (MAISTEP) is a tool which combines Random Forest  \citep[RF;][]{breiman2001random}, eXtremely randomized Trees/eXtra Trees \citep[XT;][]{wehenkel2006ensembles}, eXtreme Gradient Boosting \citep[XGBoost;][]{chen2016xgboost}, and Categorical Boosting \citep[CatBoost;][]{prokhorenkova2018catboost} algorithms, trained on a grid of stellar models to predict the radius, mass, and age of stars using atmospheric constraints. The predictions from the four algorithms are combined using a weighted sum to produce an optimal prediction. Combining  algorithms with diverse architectures increases computational demand and adds complexity, but it contributes to building a generalised highly accurate predictor. This approach known as stacking generalization \citep[][]{wolpert1992stacked,breiman1996stacked}, allows different algorithms to capture different patterns in the data, resulting in improved performance on unseen data compared to using a single algorithm. Fig.~\ref{MAISTEP} highlights the vital sections of MAISTEP and their details are summarized in Sec.~\ref{training_test_data} - \ref{stacking}.
  
\subsection{Training and test data}
\label{training_test_data}

The data for training and testing the algorithms in MAISTEP was generated using the open-source one-dimensional stellar evolution code MESA \citep[Modules for Experiments in Stellar Astrophysics:][]{paxton2018modules,paxton2019modules} version  r12778. The parameter space of the stellar grid spans a range in initial mass, [0.7 - 1.6] M$_{\odot}$ in steps of 0.05 M$_{\odot}$, initial metallicity, [Fe/H], [-0.5 - 0.5] dex in steps of 0.1, and enrichment ratios, $\Delta Y/\Delta Z$, [0.4 - 2.4] in steps of 0.4. The initial helium mass fraction, $Y_{i}$ was calculated following 
\begin{equation}
    Y_{i} = \frac{\Delta Y}{\Delta Z}Z_{i} + Y_{o}~,
\end{equation}
where, $Z_{i}$ is the initial metal mass fraction and the offset $Y_{o} = 0.2484$, is the primordial helium mass fraction \citep{cyburt2003primordial}. The grid consists of tracks evolved from zero-age main sequence (ZAMS) to end of main-sequence (defined to be a point when the central hydrogen mass fraction drops below $10^{-3}$). Microscopic diffusion was included considering only the gravitational settling component for models that do not show an unrealistic over-depletion of surface elements, and was therefore applied only to models with a ZAMS mass below $\lesssim 1.2$ M$_{\odot}$. In addition, for models with convective cores, an exponential overshoot scheme was implemented with a fixed efficiency value of 0.01. We assume that our target stars are slow rotators, and thus, the effects of rotation were neglected. Our grid employs a single solar-calibrated value of $\alpha_{\textbf{MLT}}$ = 1.71. The impact of using a single mixing length parameter in stellar grids on the inferred stellar properties has been explored in \cite{silva2015ages}. 
Table.\ref{global_physics} summarises the main aspects of our grid of stellar models. For a detailed description of the global physical inputs adopted, we refer the reader to \cite{nsamba2025uniform}.  

\begin{table}[t]
\caption{Summary of the main components of our grid of stellar models.}
    \centering
    \begin{tabular}{lc}
    \hline\hline
    Parameter & Range \\
    \hline
    Initial mass in M$_{\odot}$  & [0.7 - 1.6], $\Delta M = 0.05$\\
    Initial metallicity in dex & [-0.5 - 0.5], $\Delta$[Fe/H] = 0.1\\
    Enrichment ratio, $\Delta Y/\Delta Z$ & [0.4 - 2.4], $\Delta(\Delta Y/\Delta Z) = 0.4$\\
    \hline
    \multicolumn{2}{c}{}\\
    \hline
        Input physics & Prescription \\
        \hline
        Nuclear reaction rates  & 1, 2\\
        Solar chemical mixtures & 3\\
        OPAL equation of state & 4\\
        OPAL opacities  &  5, 6\\
        Atmosphere model & 7\\
        Atomic diffusion  & 8\\
        Core overshooting, $f_{\text{ov}} = 0.01$ & 9\\
        $\alpha_{\text{MLT}}= 1.71$ & 10\\
        \hline
    \end{tabular}
   \tablebib{ (1)~\citet{angulo1999compilation};  (2) \citet{imbriani2005s};  (3) \citet{asplund2009chemical};  (4) \citet{rogers2002updated};  (5) \citet{iglesias1996updated};  (6) \citet{ferguson2005low};  (7) \citet{krishna1966profiles};  (8) \citet{thoul1993element};  (9) \citet{herwig2000evolution};  (10) \citet{cox1968principles}.}
    \label{global_physics}
\end{table}

\subsection{Data pre-processing}
\label{preprocessing}

We selected 50 models that are nearly evenly spaced in the central hydrogen mass fraction along each evolutionary track. We follow the same technique as \cite{bellinger2016fundamental} to set up and solve an optimisation problem using \texttt{cvxpy} library \citep{boyd2004convex} in Python. This involves finding a subset of models along each track that match a desired set of evenly spaced values in the core hydrogen mass fraction. This approach ensures a well-balanced and representative sample of stellar models during the main-sequence phase.
  
The dataset is randomly split into 80\% for training and 20\% for testing our algorithms. While the training set is passed to each algorithm, the testing set is held back until the final model evaluation phase. We note that our training features only take into account  $T_{\text{eff}}$, [Fe/H], and $L$.

\subsection{Model development}
\label{model_development}
The training set is further split by employing the  $k$-fold cross-validation \citep{hastie2009elements} approach.  Each algorithm (i.e., RF, XT, CatBoost, and XGBoost) trains on $k-1$ folds (subsets) and then makes predictions for the validation set. The process is repeated $k$ times with each fold (subset) being used as a validation set once. In our case, we consider $k=$ 10.

We employ \texttt{Optuna} \citep{akiba2019optuna} to auto-tune the hyperparameters (i.e., parameters that control the structure and learning process of a model) of each base algorithm. \texttt{Optuna} incorporates cross-validation  and  explores various  hyperparameter configurations to identify the ones that deliver the best model performance, as measured by a defined metric. The library includes an in-built class to sample the user-defined parameter space, along with a pruning algorithm that monitors the intermediate result of each trial and terminates the unpromising ones, thereby accelerating the exploration process.

Given the large number of hyperparameters associated with our base algorithms, we initially used \texttt{Optuna} to determine which hyperparameters most significantly influence each algorithm's ability to predict stellar radius, mass, and age.  We then focus on tuning these key hyperparameters to streamline the hyperparameter search while maintaining each algorithm's predictive performance. 

The mean squared error metric is employed in \texttt{Optuna} to evaluate the performance of a model and  guide the optimisation process in selecting the optimal set of hyperparameters. A maximum of 50 trials are specified to balance computational efficiency while allowing thorough exploration of the hyperparameter space. The selected optimal hyperparameters (see Table~\ref{hp}) are then used to generate cross-validation predictions which served as training input (meta-features) for the non-negativity constrained least squares (nnls) optimizer. For a detailed description of each base algorithm, refer to \citet[]{breiman2001random,wehenkel2006ensembles,chen2016xgboost,breiman2017classification,prokhorenkova2018catboost}.

\begin{table*}
\caption{Optimal hyperparameters (hp) returned by \texttt{Optuna} for the radius, mass, and age by each base model.}
    \centering
    \begin{tabular}{lcccccccccccc}
        \hline\hline
        hp  & \multicolumn{3}{c}{RF} & \multicolumn{3}{c}{XT}& \multicolumn{3}{c}{XGBoost}& \multicolumn{3}{c}{CatBoost} \\
         & Radius & Mass & Age & Radius & Mass & Age &Radius & Mass & Age &Radius & Mass & Age \\
        \hline
        learning\_rate  & - & - & -  & - & -  & - &0.29&0.3 &0.3 &0.64 &0.97  &0.95\\
        depth /   &20&20&20 &20  &20  &20 &6&6 &6 &5  &5  & 5 \\
        \hspace{0.2cm} max\_depth  & &  &  &  &  & && & &&&\\
        n\_estimators /  &300&224&75 & & &  &1000  & 1000 & 1000 &1438 & 1172  &1495\\
        \hspace{0.3cm}iterations  & &  &  &  &  & && & &&&\\
        \hline
    \end{tabular}
    \tablefoot{def signifies that the  default value was considered. Default values were also used for any hp that are excluded here.
    }
    \label{hp}

\end{table*}

\subsection{Stacking approach}
\label{stacking}
The predictions from the base algorithms, generated during cross-validation, are assigned non-negative weights, with the most accurate predictions typically receiving higher weights. The weights are determined by the least squares optimisation method \citep{lawson1995solving}, which adjusts them to minimize the overall prediction error. Essentially, given the base algorithms labelled, $j = 1, 2, 3, 4$, each making predictions $p_{ij}$, for the $i^{th}$ data point in a sample of size $n$, we seek to find the weights $\beta_{j}$, such that;

\begin{equation}
    \text{min} \sum_{i=1}^{n} (\hat{y}_i - y_i)^2~,
\end{equation}
with
\begin{equation}
  \hat{y}_i = \sum_{j=1}^{k}\beta_{j}.p_{ij}~,
  \label{pred}
\end{equation}
subject to $\sum \beta_{j} = 1$,  $\beta_{j} \geq 0$. $y_{i}$ is the actual value of the $i^{th}$ data point and $\hat{y}_{i}$ is the associated weighted prediction defined by equation \ref{pred}. This approach allows for a clear assessment of how much each base algorithm contributes to the final stacked model's prediction, with the weights directly reflecting each algorithm's relative importance. Since the weights are constrained to be non-negative and sum to 1, the interpretation is straight forward: higher weights indicate a greater contribution to the final prediction. The optimal weights generated are adopted for producing the final predictions during the test and the application phase, via the stacking approach. We import the nnls optimizer from the \texttt{scipy.optimize} library into  our Python code.  

\section{Evaluation of MAISTEP}
\label{maistep_eval}
We conduct a double assessment of our code using both artificial data from stellar models and real observed stars. We utilise the root mean squared error (RMSE) to evaluate and compare the performance of the base algorithms and our weight-based model on test data. We choose RMSE because it retains the same unit as the target variable, making it easy to interpret the bias.

\subsection{Performance on artificial data}
\label{artificial_data}
\begin{figure}[t]
        \centering
        \includegraphics[scale=0.45]{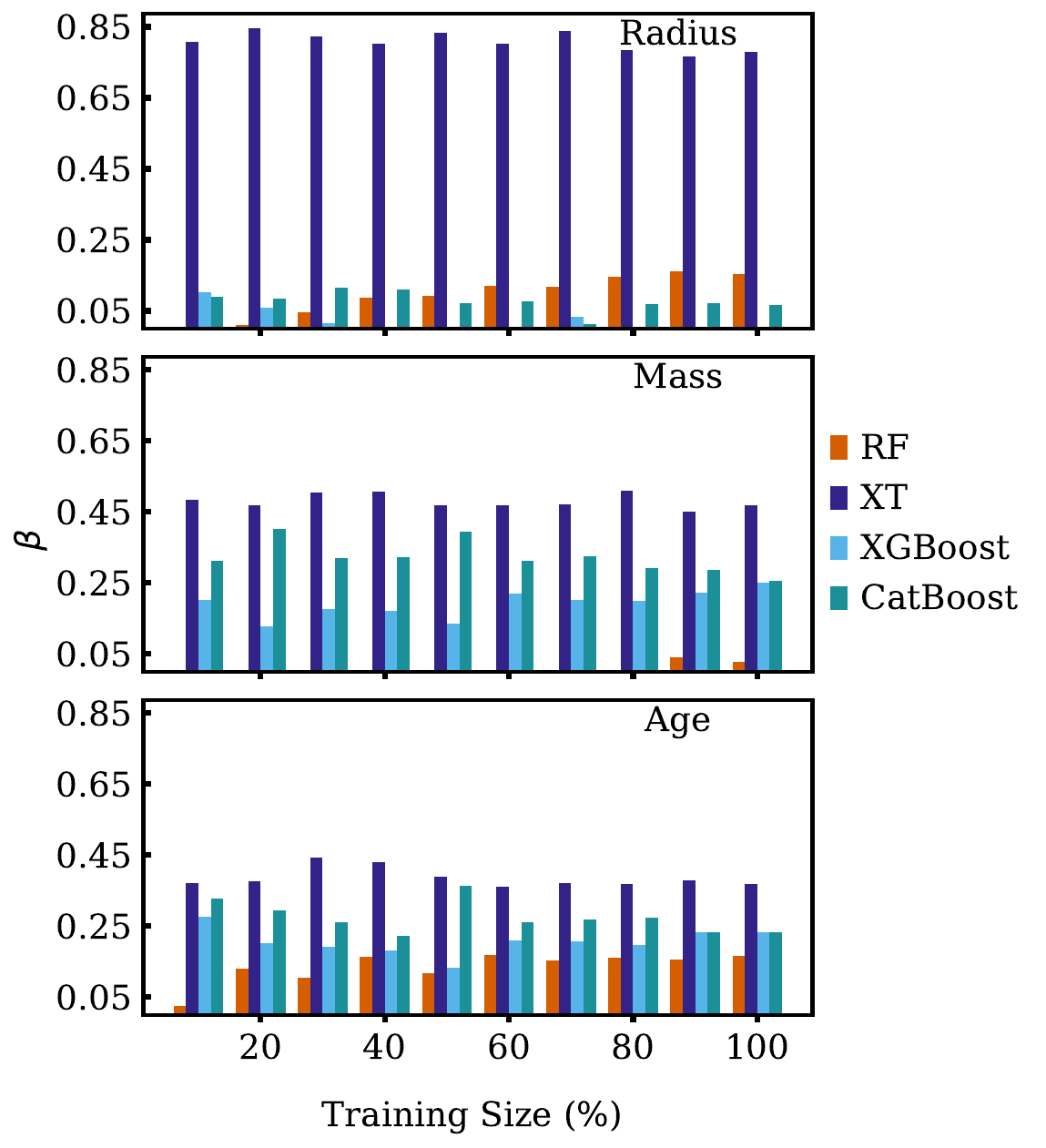}
            \caption{Distribution of the weights ($\beta$) derived using nnls optimizer, which are employed to combine predictions of stellar radius (top panel), mass (middle panel), and age (bottom panel) from the base algorithms. See text for details.}
            \label{weights}
\end{figure}

  We recall that our data is divided into 80\% training and 20\% test sets. We further partitioned the training set into subsets ranging from 10\% to 100\% in increments of 10\%. For each subset, optimal weights used to stack the predictions generated through cross-validation (see Sec.~\ref{model_development}) by the base algorithms were determined using nnls, as described in Sec.~\ref{stacking}. These weights are shown in Fig.~\ref{weights}. Overall, we observe that higher weights are consistently assigned to XT predictions in estimating the radius, mass, and age, across all training sizes. This suggests that XT's predictions are generally more accurate compared to those from other algorithms, at least within the scope of our tuned hyperparameters. Furthermore, while variations in weights are observed at different training sizes, they tend to stabilize as training data approach full size, except in the case of mass. 
  
  To evaluate the generalisation capability of our base algorithms, we trained each algorithm on each subset. We then made predictions for the 20\% test set. This evaluation assumes that the input features (i.e., $T_{\text{eff}}$, [Fe/H], and $L$) used to predict the radius, mass, and age of the test set are free from errors and bias. Finally, we compared the performance of each base model on the test set to that of the stacking model.
 \begin{figure}[t]
        \centering
        \includegraphics[scale=0.3]{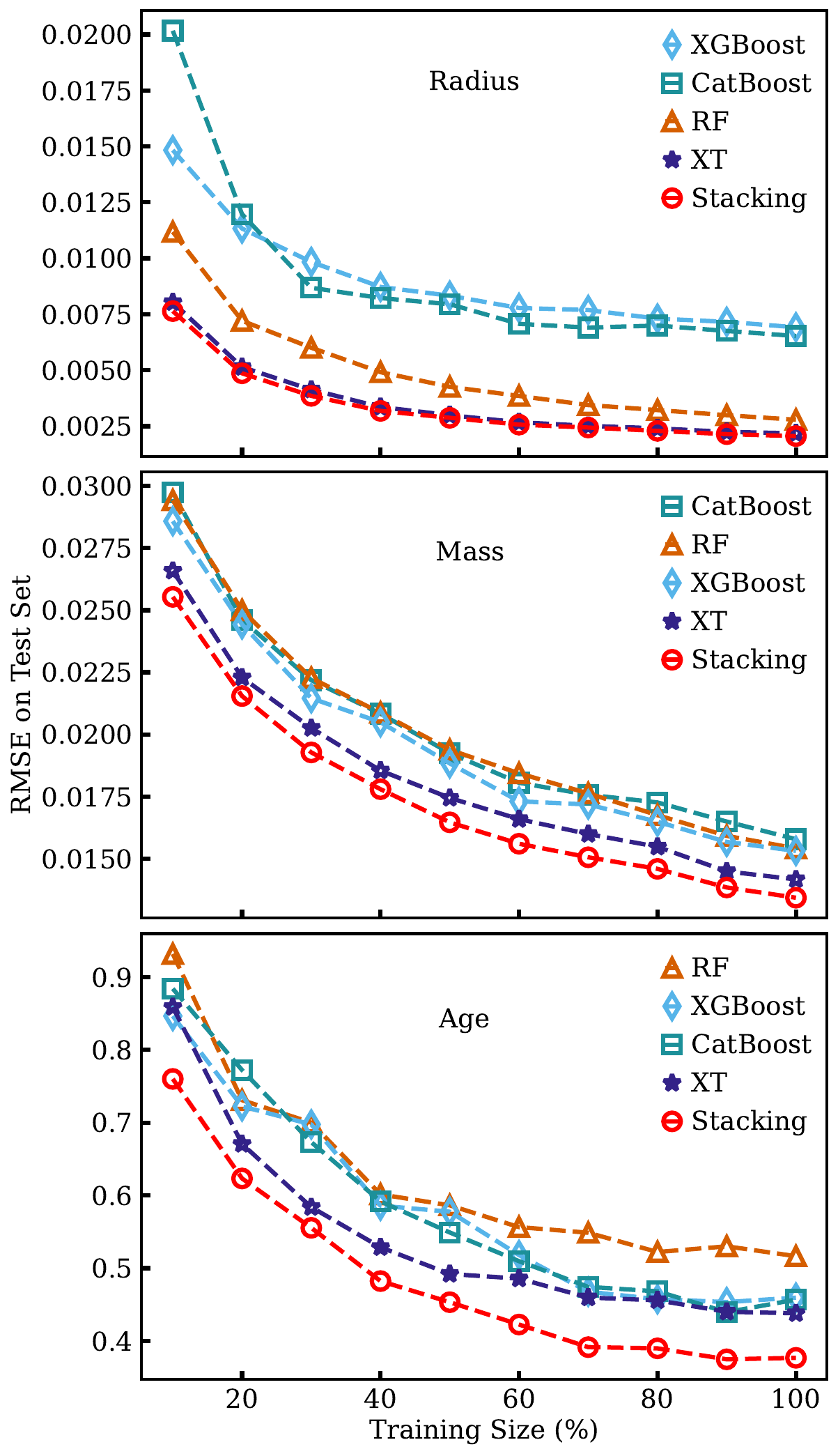}
            \caption{Evaluation metric on the test sample as a function of the training size. The legend is organized in the order of performance, with the worst performers at the top and the best  performers at the bottom.}
            \label{radius_mass_age_rmse}
\end{figure}

  Figure.~\ref{radius_mass_age_rmse} shows that as more training examples become available, each algorithm  learns and improves its ability to predict the radius, mass, and age, resulting into a reduced  bias on the test set. For radius (top panel of Fig.~\ref{radius_mass_age_rmse}), the learning continues until a stable regime (saturation point) is reached with about 50\% of the training samples, beyond which no additional improvement occurs, within the framework of our hyperparameter tuning. The early attainment of the saturation point could be due to two of the training features (i.e., $T_{\text{eff}}$ and $L$) directly constraining the radius through the Stefan-Boltzmann relation \citep{planck1900theory}.  In addition, we note that the stellar model luminosity, $L$ is derived using the same relation with the $T_{\text{eff}}$ and $R$ as inputs, in stellar evolutionary codes. Therefore, these findings may be consistent due to the non-independence of $L$, yet they still demonstrate that an accurate $L$ coupled with $T_{\text{eff}}$, can effectively recover the stellar radius using any of our ML algorithms.

  In contrast, the RMSE curves for the mass (middle panel of Fig.~\ref{radius_mass_age_rmse}) show that learning continues up to 100\% training size, while for the age (bottom panel of Fig.~\ref{radius_mass_age_rmse}), they flatten out at $\sim 90$\%. In particular, for the mass, this could suggest that the algorithms required more samples (information) to explore in order to achieve adequate training. To test this, we trained our algorithms on the entire dataset and made predictions for some observed stars through stacking. We find that increasing the training sample had almost no influence on the resulting stellar masses.
  
  Overall, within the scope of our hyperparameter tuning, XT stands out as the best performing base algorithm for our regression tasks in predicting stellar radius, mass, and age.  It is also weighted the most in our stacking approach across all training sizes as seen in Fig.~\ref{weights}. XT's good performance likely stems from its highly randomized approach to feature selection and choosing split points, which allows  for an effective balance between the bias and variance while maintaining a competitive accuracy. Therefore, we recommend it for use when computational resources are limited \citep[we refer the reader to][for a full exploration of XT]{bellinger2016fundamental}. However, we observe enhanced performance when XT is combined with the other three algorithms, particularly in reducing the bias in age predictions, resulting into an increase in accuracy of $\sim$ 7\% (compared to XT) at full training size (see Fig.~\ref{radius_mass_age_rmse}). Therefore, our stacking approach demonstrates a distinct advantage in performance, as it effectively combines the strength of the base algorithms, making it the preferred method of choice for inferring the radius, mass, and age of observed stars.

\subsection{Performance on real stars}
\label{real_data}
We apply MAISTEP in its stacking approach setting, along with the associated weights to characterize real stars in the APOKASC \citep{serenelli2017first} and LEGACY  \citep[][]{silva2017standing,lund2017standing} samples. We select targets that fall within the parameter space of the stellar models employed in the training process. We pass the observations in $T_{\text{eff}}$, [Fe/H], and $L$, to predict radius, mass, and age. To check for consistency, we also compare the predicted stellar properties to those obtained in the two data sources (APOKASC and LEGACY sample) which employ different methods and incorporate both seismic and spectroscopic constraints in their optimisation processes. 

The APOKASC  catalogue  consists of 415 (main-sequence and sub-giant) stars whose properties were determined utilising global asteroseismic observables (i.e. large frequency separation, $\Delta\nu$ and frequencies at maximum power, $\nu_{\text{max}}$) complemented with effective temperatures, $T_{\text{eff}}$ and metallicities, [Fe/H]. Of the targets, 167 have stellar properties within the parameter space of our training grid. This excludes stars that overlap with the LEGACY sample, since they are accounted for in the LEGACY comparisons. In addition, we also exclude stars identified as binaries according to SIMBAD and/or with astrometric solution above the favoured renormalised unit weight error (RUWE) threshold of $\lesssim 1.4$ in $Gaia$ DR3\footnote{https://gea.esac.esa.int/archive/} \citep{lindegren2018gaia}.  
\cite{serenelli2017first} present two sets of results based on different temperature scales. We compare our findings to those obtained using the photometric $T_{\text{eff}}$ scale from the SDSS $griz$ bands, as recommended by the authors.

The LEGACY sample is composed of 66 main-sequence $Kepler$ targets \citep[][]{silva2017standing,lund2017standing}. The stellar radii, masses, and ages were determined using either individual mode frequencies or frequency ratios together with 
 spectroscopic effective temperature and metallicity compiled from different literature sources \citep[i.e.][]{ramirez2009accurate,pinsonneault2012revised,huber2013fundamental,pinsonneault2014apokasc,chaplin2013asteroseismic,casagrande2014towards,buchhave2015metallicities}. \cite{silva2017standing} demonstrates a disagreement between parallaxes obtained from asteroseismic distances (which they computed from a combination of Infra-red flux measurements of angular diameter and asteroseismic radius) and $Gaia$ parallaxes, with discrepancies exceeding 3$\sigma$ for some of the stellar targets. They emphasize that the flux of those stars could be contaminated by their companions, and since that directly impacts the luminosity determinations, we exclude them. These include: KIC 9025370, KIC 7510397, KIC 8379927, KIC 10454113, KIC 9139151, KIC 9965715, KIC 7940546, KIC 4914923, KIC 12317678. Furthermore, we exclude KIC 10068307 and the spectroscopic binary component KIC 6933899. Hence, a total of 55 stars from the LEGACY sample are considered in our analysis.

 \subsubsection{Luminosity determination}
 \label{lum}
\begin{figure}
        \centering
        \includegraphics[scale=0.4]{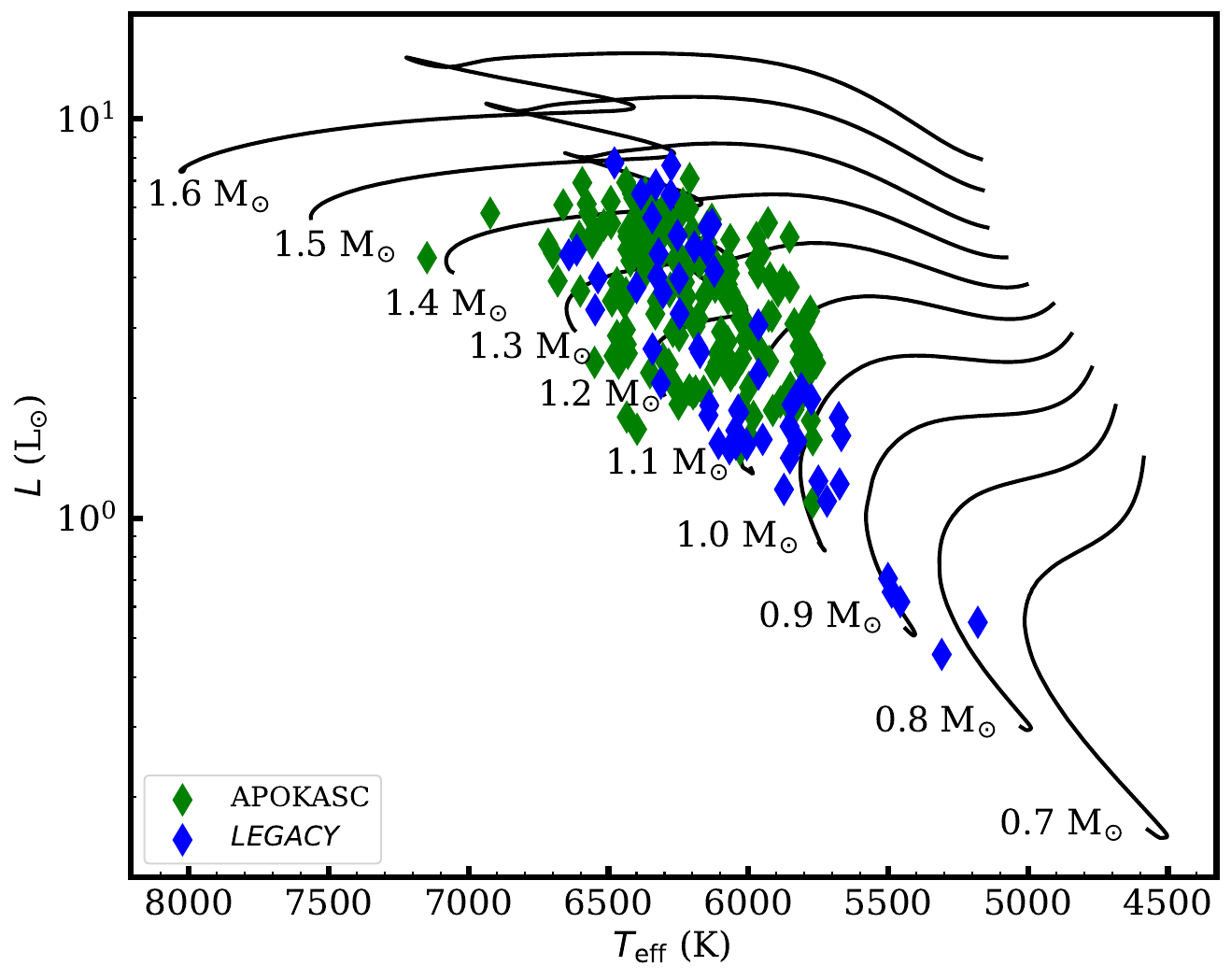}
            \caption{HR diagram showing our selected sample, which include 167 stars from APOKASC catalogue plotted in green and  55 stars from the LEGACY sample shown in blue diamonds. In addition, we over-plot a few tracks at solar metallicity, [Fe/H] = 0, extending up to the bottom of the red-giant branch.}
        \label{hr}
\end{figure}

 Both the APOKASC and LEGACY data sources provide measurements of metallicities and effective temperatures. We take advantage of the precise parallaxes from $Gaia$ data in DR3 and follow \cite{pijpers2003selection} to determine the corresponding luminosities using the expression
 \begin{equation}
 L = 10^{4.0 + 0.4M_{\text{bol},\odot} - 2\text{log}\pi (\text{mas}) -0.4(m_{\text{g}} - A_{\text{g}} + \text{BC}_{\text{g}})}~,
\label{lum_eqn}
\end{equation}
where, $\pi$, and $m_{\text{g}}$ represent $Gaia$ parallax and apparent g-band magnitudes \citep{collaboration2022gaia}, respectively. The bolometric corrections $\text{BC}_{\text{g}}$ for the sample were computed using coefficients from \cite{andrae2018gaia}, and we adopted $M_{\text{bol},\odot} = 4.74$ as recommended by \cite{mamajek2015iau}. We correct for the effects of interstellar absorption by computing the extinction, defined by

\begin{equation}
   A_{\text{g}} = R_{\text{g}}\times E(B - V)~, 
\end{equation}
 where $E(B - V)$ is the color excess which was queried using 3D dustmaps\footnote{https://dustmaps.readthedocs.io/en/latest/} python package \citep{2018JOSS....3..695M}. We adopt an extinction coefficient, $R_{\text{g}} = 3.303$, according to \citep{schlafly2011measuring}. 
The dust maps do not provide uncertainties on the reddening.  Therefore, to compute lower and upper uncertainty levels on luminosity, we employ standard error propagation considering only the uncertainty in parallax and apparent magnitude. In Fig.~\ref{hr}, we display the approximate positions of our selected targets on the HR diagram with evolutionary tracks constructed at solar metallicity running from ZAMS to the bottom of the red-giant branch.

To propagate the uncertainties in $T_{\text{eff}}$, [Fe/H], and $L$ to our inferred properties, we randomly generate 10000 realisations  based on the standard deviations of these observables. These realisations are then input into the trained algorithms, resulting in 10000 predictions for the radius, mass, and age of each star. From the distributions, we compute the median and quartiles.

\subsubsection{Comparison with APOKASC}
\begin{figure}
        \centering
        \includegraphics[scale=0.32]{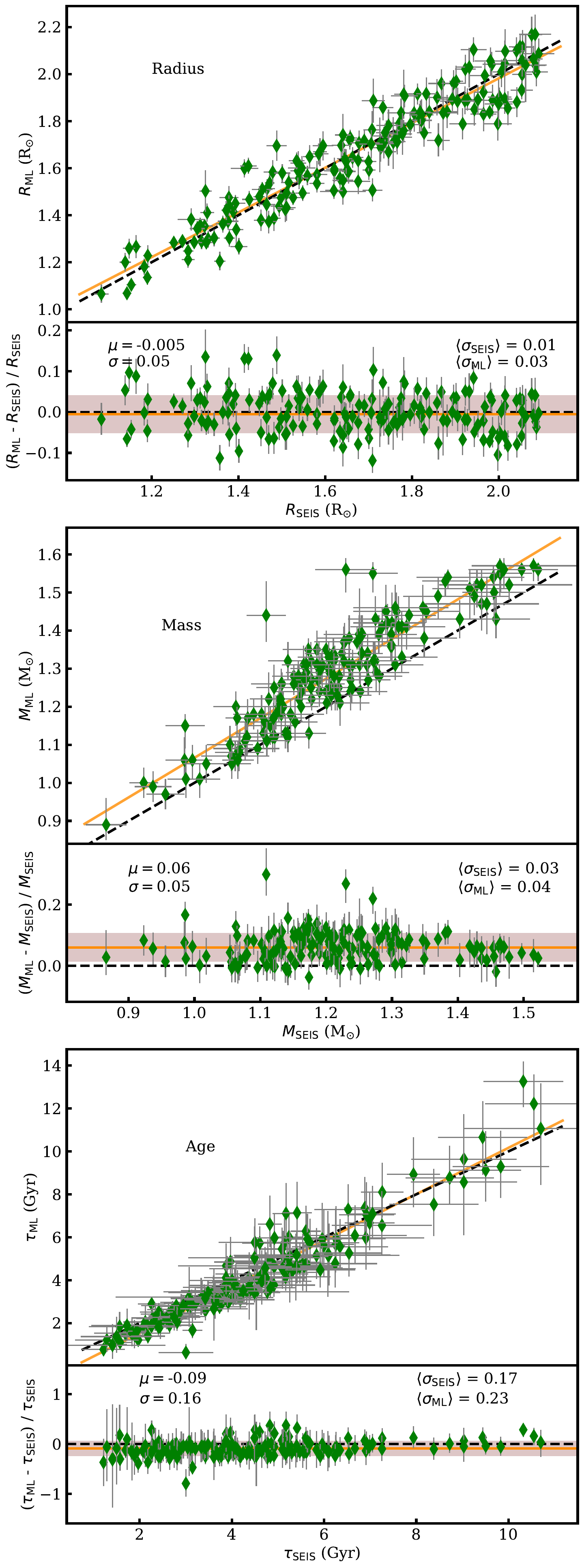}
            \caption{Each panel contains a one-to-one relation and fractional differences from MAISTEP with respect to APOKASC as the reference values, for radius (top), mass (middle), and age (bottom).
            The black dashed lines in all panels represent the unity relation.
            The orange line in the one-to-one plots indicate the best-fit line while in the fractional difference plots, it represents the mean offset/bias, $\mu$. The brown shaded region highlights the associated scatter, $\pm\sigma$. Additionally, we give the average statistical relative uncertainties in angle brackets.}
            \label{APOKASC_radius_mass_age}
\end{figure}

In Fig.~\ref{APOKASC_radius_mass_age}, we compare the estimates of the stellar radii, masses, and ages obtained using MAISTEP to those in APOKASC. We find a good agreement between the MAISTEP radius values, $R_{\text{ML}}$, with those from APOKASC, $R_{\text{SEIS}}$ (see top panel of Fig.~\ref{APOKASC_radius_mass_age}). We observe a scatter of 5\% and an offset/bias of -0.5\%. The average  relative uncertainty in radius from our tool is 3\%, which is higher than the 1\% uncertainty in the APKOSAC radius estimations.

\begin{figure}
        \centering
        \includegraphics[scale=0.32]{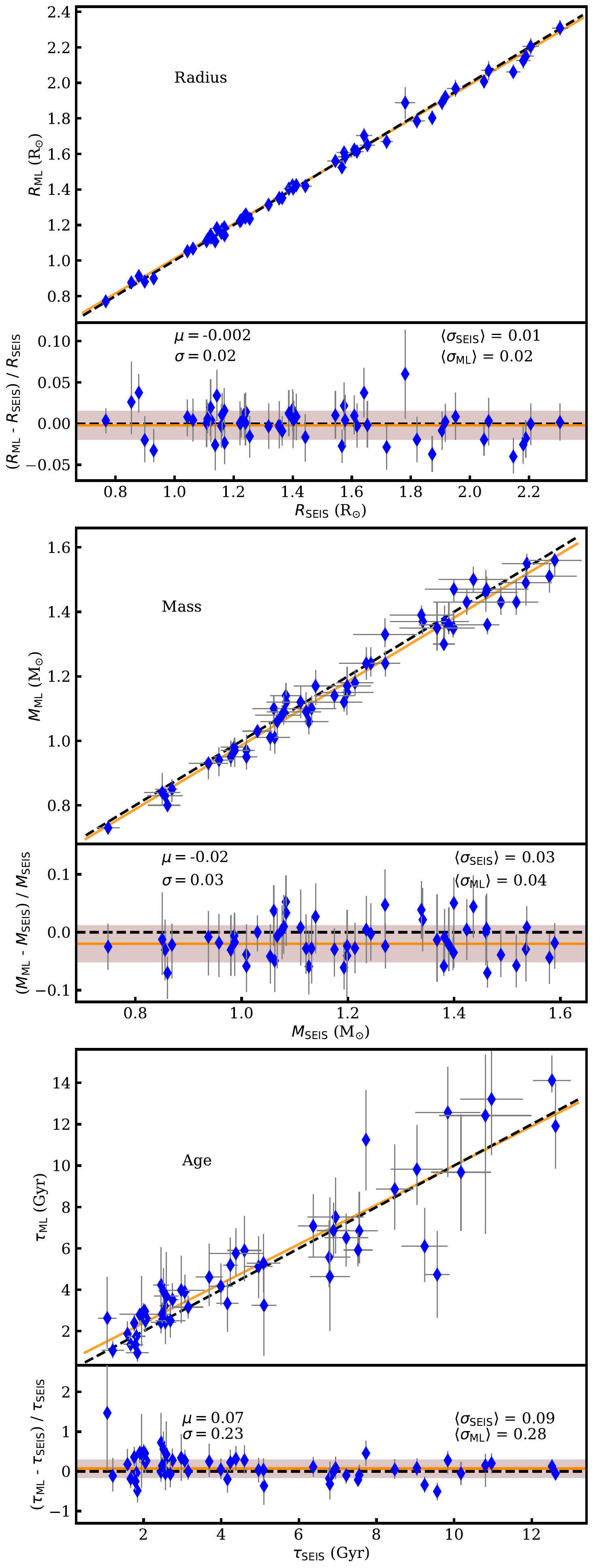}
            \caption{Same as Fig.~\ref{APOKASC_radius_mass_age}, but comparing estimates from MAISTEP and AIMS for the stars in the LEGACY sample.
            }
            \label{LEGACY_radius_mass_age}
\end{figure}

The middle panel of Fig.~\ref{APOKASC_radius_mass_age} shows that the masses inferred using MAISTEP are systematically higher by 6\%, and we obtain a scatter of 5\%.  The average statistical relative uncertainty in mass from MAISTEP is 4\%, compared to 3\% from APOKASC. Furthermore, this translates into MAISTEP yielding systematically lower ages with a mean offset of -9\% and a scatter of 16\% (see the bottom panel of Fig.~\ref{APOKASC_radius_mass_age}). The average relative uncertainty in age is 23\% and higher than the 17\% based on estimations in the APOKASC. The observed offsets could stem from variations in the model physics and the sets of constraints employed.

\subsubsection{Comparison with LEGACY}
\cite{silva2017standing} provides seven sets of results generated by different optimisation codes adopting grids with different stellar model physics. 
Here, we compare our results with those inferred using the AIMS \citep[Asteroseismic Inference on a Massive Scale:][]{rendle2019aims} pipeline. AIMS infers stellar radii, masses, or ages by combining the likelihood function with prior distributions on those properties, followed by an exploration of parameter space using MCMC to obtain the resulting distributions. The grid of models adopted in AIMS by \citet{silva2017standing} was constructed by \cite{coelho2015test} using the MESA stellar evolutionary code, without taking into account atomic diffusion. In the optimisation process, individual oscillation mode frequencies were complemented with spectroscopic effective temperatures and metallicities. 

The top panel of Fig.~\ref{LEGACY_radius_mass_age} presents a comparison between the  radius estimates obtained using MAISTEP and AIMS. We find a bias of -0.2\% and a scatter of 2\%, demonstrating an excellent agreement in radius determinations by the two approaches. The statistical average relative uncertainties in radius from MAISTEP and AIMS are consistent, i.e. 2\% and 1\%, respectively. In the middle panel of Fig.~\ref{LEGACY_radius_mass_age}, we observe a bias  of -2\%  and a scatter of 3\% in mass with MAISTEP estimations being systematically lower than those from AIMS.  The average relative uncertainty in mass for the MAISTEP and AIMS estimations is 4\% and 3\%, respectively. For the age, estimated obtained using MAISTEP are systematically higher by 7\% and a scatter of 23\% (see bottom panel of Fig.~\ref{LEGACY_radius_mass_age}). ML estimates also have an average relative age uncertainty of 28\%, compared to 9\% from AIMS. The systematically higher ages inferred by MAISTEP contrast with the expected age estimates from models with and without atomic diffusion. The observed discrepancy may be explained by the systematically lower masses inferred by MAISTEP, which could arise from the differences in the observational constraints used in MAISTEP compared to the LEGACY analyses. Overall, MAISTEP predicts stellar radii, masses, and ages that are commensurate with seismic inferences.  

\section{Ages of giant-planet hosts}
\label{giant_planets}
Over the past two decades, the discovery of exoplanets has consistently defied our expectations set by the solar system, uncovering a much greater diversity of planetary systems than ever thought. One of the most remarkable discoveries is the existence of close-in Jupiter-mass planets (the Hot Jupiters), with orbital periods of just a few days \citep{mayor1995jupiter,rasio1996dynamical}, which has continually raised questions about their formation and longevity. More recently, with the substantial amount of exoplanet detections and our growing understanding of the star-planet connections, some statistical studies have focused on analysing the ages of the host stars to gain better insights into the evolution of the Hot Jupiters. For instance, \cite{miyazaki2023evidence} used ages inferred from stellar models (isochrones) for a sample of 382 Sun-like  stars, which includes 124 giant planets, from the California Legacy Survey catalogue to demonstrate that Hot Jupiters are preferentially
hosted by relatively younger stars, and that their number decreases with stellar age. \cite{hamer2019hot} and \cite{chen2023evolution} illustrate their findings using age proxies, namely, Galactic space velocities and the average age-velocity dispersion relation, respectively, and reach similar conclusions. The explanation provided is that tidal interactions between Hot Jupiters and their host stars during the main-sequence phase lead to orbital decay over time, causing these planets to spiral inward and ultimately be engulfed. As a result, the number of Hot Jupiters decreases at the later stages of the main-sequence, suggesting that they are typically found around young stellar hosts.  We aim to apply MAISTEP to infer the ages of individual stars from a relatively large sample consisting solely of planet hosts, in order to test the above claims.

 We collect planet data from the NASA Exoplanet Archive\footnote{https://exoplanetarchive.ipac.caltech.edu/} and the Extrasolar Planets Encyclopaedia\footnote{http://www.exoplanet.eu/}. We select stars orbited by giant planets with the best available mass estimate in the following order of preference: $M_{\text{P}}$, $M_{\text{P}}\times$sin$i$/sin$i$, or $M_{\text{P}}\times$sin$i$, depending on data availability. The selected planets span a range in Jupiter mass of 0.3 to 13 \citep[13$M_{\text{J}}$ is the lower limiting  mass of brown dwarfs;][]{des2022iau}. We identify Hot Jupiters as giant planets orbiting very close to their host stars with  periods of $\leq$ 10 days. Our control sample consists of stars hosting the Warm and Cold Jupiters, with the same mass range as the Hot Jupiters. The Warm Jupiters have orbital periods of $ 10 < P (\text{days}) \leq 365$, while the Cold Jupiters have orbital periods of $ 1 < P (\text{years}) \leq 10$ \citep{miyazaki2023evidence}.  We then cross-match the hosts that have confirmed Jupiter-mass planets with SWEET-Cat\footnote{https://sweetcat.iastro.pt/}, selecting only stars with homogeneously derived spectroscopic $T_{\text{eff}}$ and [Fe/H] \citep{santos2013sweet}. 

The stellar luminosities were derived using the equation \ref{lum_eqn}. We select main-sequence stars with observable properties within the parameter space of ML training grid. Our sample consists of 427 stars hosting 470 giant planets, including 208 Hot Jupiters, 96 Warm Jupiters, and 152 Cold Jupiters. Of these stars, 387 host single planets, 37 host double planets, and 3 host three planets. We clarify that, unless stated otherwise, for stars hosting multiple planets, if the planets share a similar range in orbital periods, the star is counted only once. However, if the planets have orbital periods that place them in different categories, the star is included in each of those categories. 

\begin{figure}[t]
    \centering
    \includegraphics[scale=0.48]{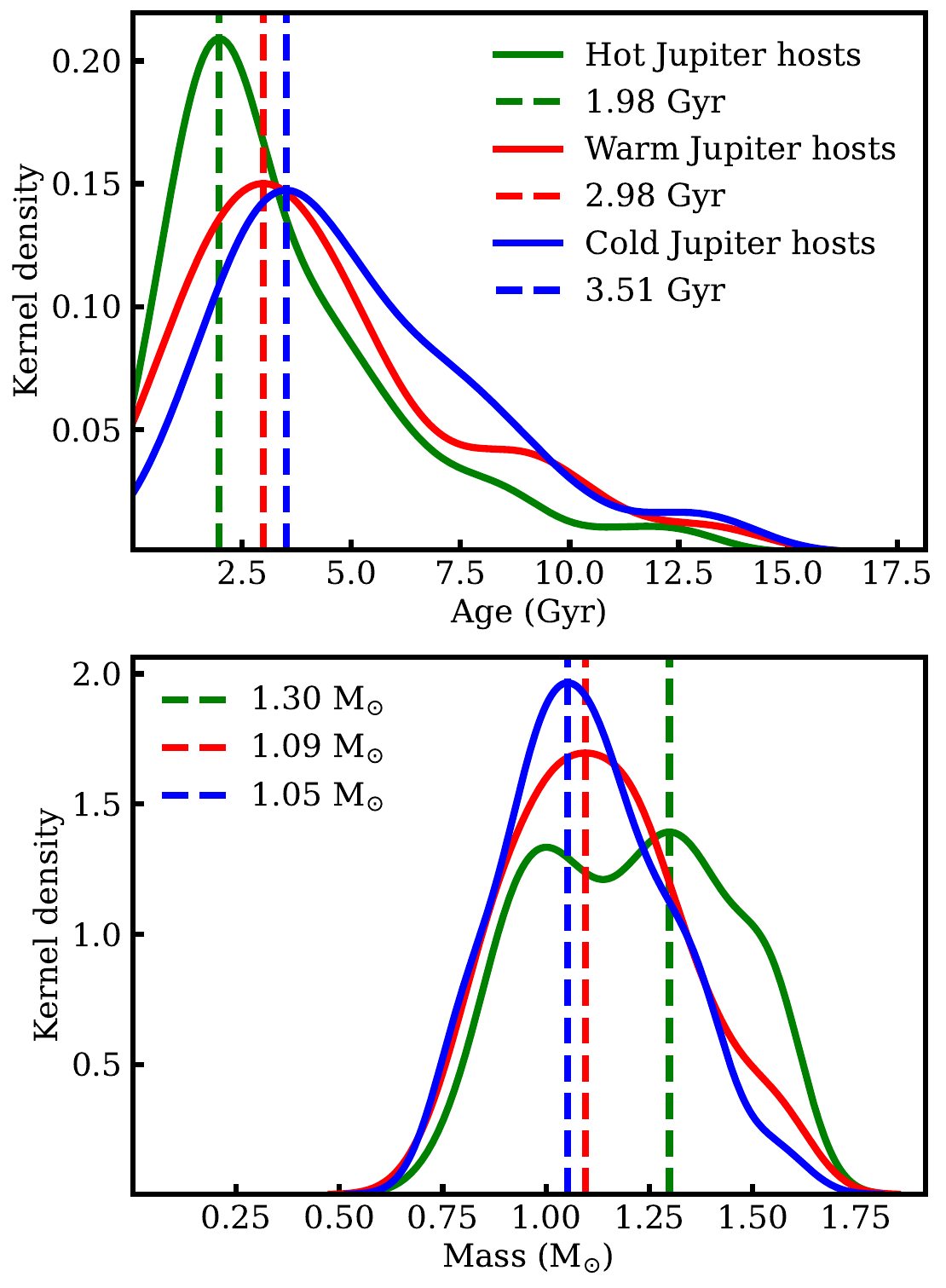}
    \caption{Kernel density illustrating the age (top panel) and mass (lower panel) distributions of planet hosting stars.}
    \label{planet_host_ages_masses}
\end{figure}

Following the same procedures as discussed in Sec.~\ref{real_data}, we calculate the ages of the giant-planet hosts. In the top panel of Fig.~\ref{planet_host_ages_masses}, we show the resulting age distributions of stellar populations hosting Hot Jupiters, Warm Jupiters, and Cold Jupiters, which are constructed solely from the median ages. The data show distinct peaks at ages of 1.98, 2.98, and 3.51 Gyr, with corresponding mean ages of 3.52, 4.41, and 5.13 Gyr, respectively. 
 
\begin{table*}[h!]
\caption{Statistical test results from the age and mass distributions.}
    \centering
    \begin{tabular}{lcccccc}
        \hline\hline
        Category  & \multicolumn{2}{c}{AD test} & \multicolumn{2}{c}{KS test} & \multicolumn{2}{c}{t-test} \\
         & AD & \textit{p}-value & KS & \textit{p}-value  & t & \textit{p}-value \\
         \hline
         \multicolumn{7}{c}{Age}\\
        \hline
        A  & 7.81 & $< 10^{-3}$ & 0.23 & $8\times10^{-4}$  & 3.39 & $8\times10^{-4}$ \\
        B  &30.5 & $< 10^{-3}$ & 0.36 &$3.07\times10^{-11}$   & 6.28 & $1.17\times10^{-9}$  \\
        C  & 2.74 & 0.025 & 0.16 & 0.07  & 1.57 & 0.12\\
        \hline
        \multicolumn{7}{c}{Mass}\\
        \hline
        A  & 4.4 & 0.006 & 0.19 & 0.01  & 2.9 & 0.004 \\
        B  &11.6 & $< 10^{-3}$ & 0.36 &$1.0\times10^{-4}$   & 4.6 & $5.8\times10^{-6}$  \\
        C  & -0.53 & 0.25 & 0.08 & 0.81  & 0.98 & 0.33\\
        \hline
    \end{tabular}
    \label{tab:test_results}
    \tablefoot{The Anderson-Darling (AD) test, Kolmogorov-Smirnov (KS) test, and t-test for three categories: A - Hot-Jupiter hosts vs Warm-Jupiter hosts; B - Hot-Jupiter hosts vs Cold-Jupiter hosts; and C - Warm-Jupiter hosts vs Cold-Jupiter hosts.
    
    }
\end{table*}

To quantify the age differences further, we conduct statistical tests on the distributions between the categories. Specifically, we tested the null hypothesis that the distributions are identical and that the observed mean age differences are not significant. We used Anderson-Darling \citep{anderson1954test,scholz1987k} and Kolmogorov-Smirnov \citep{kolmogorov1933sulla,smirnov1939ecarts} for $k$ samples to assess the distributions, along with the t-test \citep{student1908probable} to compare the mean values. These functions were sourced from \texttt{scipy.stat} in Python 3.10.12. We summarize the results in Table.\ref{tab:test_results}, for three categories: A - Hot-Jupiter hosts versus Warm-Jupiter hosts; B - Hot-Jupiter hosts versus Cold-Jupiter hosts; and C - Warm-Jupiter hosts versus Cold-Jupiter hosts. 

We found low $p-$ values at the 5\% significance level for categories A and B, indicating that the distributions differ and the mean age differences are statistically significant. For category C, both the t and the KS tests fail to reject the null hypothesis, but the AD test rejects with a $p = 0.025$, suggesting that there may be a noticeable difference between the two distributions that the other two tests did not detect. This could be because of the AD's sensitivity to the tails of the distributions. 
In addition,  we conducted a test that excluded stars hosting planets in different categories. The results showed that the Warm and Cold Jupiter hosts are identical, with $p-$ values of 0.05, 0.22, and 0.09. Generally, the statistical tests confirm that Hot Jupiters are hosted by stars that are relatively younger than those hosting the Warm and Cold Jupiters. This is consistent with previous findings, which examined a relatively small sample of host stars \citep[][]{miyazaki2023evidence}, as well as with kinematic results \citep[e.g.,][]{hamer2019hot,chen2023evolution}. 

In the bottom panel of Fig.~\ref{planet_host_ages_masses}, our sample shows that the Hot-Jupiter hosts are statistically more massive than stars hosting the Warm and Cold Jupiters, which may help explain the observed age differences.  The dip in the mass distribution of Hot Jupiters could be a selection effect, as our study considers only hosts with homogeneously derived spectroscopic properties in the SWEET-cat catalogue. The $p-$values from the AD test, the KS and the t-test, for categories A and B, are all below the critical value of 0.05 (see Table.~\ref{tab:test_results}) suggesting that the mass distributions are not identical. For category C, all tests do not reject the null hypothesis, with $p-$values greater than 0.05. Therefore, the Warm and Cold Jupiter mass distributions and associated mean values are similar, but they are statistically different when compared with the Hot-Jupiter host masses. Further, to assess the sensitivity of the statistics to the orbital period threshold set for the Hot-Jupiter hosts, we varied the period over a range of 5 to 15 days in steps of 1 day. The results show no significant departure from the conclusions drawn based on a period threshold of $P \leq 10$ days. 

\section{Summary and conclusions}
\label{summary_conclusion}

We developed a grid-based machine learning tool (MAISTEP)  for estimating robust stellar radius, mass, and age using atmospheric constraints. 
The code relies on transfer learning in which various machine algorithms (i.e. RF, XT, CatBoost, XGBoost) pre-trained on data from stellar models are used to make predictions for real stars. Most importantly, MAISTEP, has enhanced generalisation capabilities producing highly accurate stellar radius, mass, and age, in contrast to the typical use of single algorithms. 

We explored the robustness of MAISTEP by conducting a detailed comparison with inferences from APOKASC and LEGACY samples. In terms of radius, we observe a bias of -0.5\% with a scatter of 5\% when compared to the APOKASC, and a bias of -0.2\% with a scatter of 2\% when compared to the LEGACY sample. A bias of 6\% and -2\%,  with  scatter values of 5\% and 3\%, is obtained in mass when compared to the APOKASC and LEGACY sample results, respectively. For age, the bias is -9\% and 7\%, with scatter values of 12\% and 23\% when compared to the APOKASC and LEGACY samples, respectively. Accordingly, our results demonstrate that we can infer stellar masses and radii to a precision relevant for exoplanetary studies, in cases where seismic information is not available. Our findings also emphasise the need for seismic information in order to yield robust ages. In addition, in cases where one needs to explore the internal structures of stars, asteroseismology offers a solution.

Lastly, we use MAISTEP to estimate ages of planet-hosting stars that have homogeneously derived spectroscopic effective temperatures and metallicities, to which we complement with $Gaia-$based luminosities. We find consistent results with previous finding that Hot-Jupiter planets are preferentially hosted by relatively a younger and massive stellar population compared to the Warm/Cold Jupiter hosts. 
 
Our study primarily focused on the stars in the main-sequence; however, we plan to extend it to the sub-giant and red-giant branches where there is a growing number of detections of high-mass planet-hosting stars, supported by their slow surface rotations and cooler atmospheres, both of which are advantageous for radial velocity measurements. In addition, we plan to use our tool to investigate the impact of stellar model physics on an ensemble of planet-host stars.

\begin{acknowledgements} 
\footnotesize
J.K. acknowledges funding through the Max-Planck Partnership group - SEISMIC Max-Planck-Institut für Astrophysik (MPA) -- Germany and Kyambogo University (KyU) - Uganda. 
B.N. acknowledges postdoctoral funding from the "Branco Weiss fellowship – Society
in Science" through the SEISMIC stellar interior physics group. T.L.C.~is supported by Funda\c c\~ao para a Ci\^encia e a Tecnologia (FCT) in the form of a work contract (CEECIND/00476/2018). V.A. and N.C.S. acknowledge support from Funda\c{c}\~ao para a Ci\^encia e a Tecnologia through national funds and by FEDER through COMPETE2020 - Programa Operacional Competitividade e Internacionalização by these grants: UIDB/04434/2020; UIDP/04434/2020; 2022.06962.PTDC. Funded/Co-funded by the European Union (ERC, FIERCE, 101052347). Views and opinions expressed are however those of the author(s) only and do not necessarily reflect those of the European Union or the European Research Council. Neither the European Union nor the granting authority can be held responsible for them. 
J.K. and B.N. also acknowledge funding from the UNESCO-TWAS programme, ``Seed Grant for African Principal Investigators'' financed by the German Ministry of Education and Research (BMBF).
J.K. also thanks Andreas W.N., Pedro A.C.C., Thibault B. at IA-CAUP, and Earl P.B at Yale University, for the useful discussions regarding machine learning.

$Software$: Data analysis in this manuscript was carried out using the Python  3.10.12 libraries; scikit-learn 1.5.1 \citep{pedregosa2011scikit}, pandas 2.2.2 \citep{mckinney2010proceedings}, NumPy 1.26.4 \citep{van2011numpy}, and SciPy 1.14.1 \citep{virtanen2020scipy}
 \end{acknowledgements}




\bibliographystyle{aa}
\bibliography{mybib} 



\appendix
\twocolumn
\section{Additional tests}
\subsection{Comparison with other LEGACY results}
\label{extra_legacy_results}
\begin{figure}[t]
        \centering
        \includegraphics[scale=0.33]{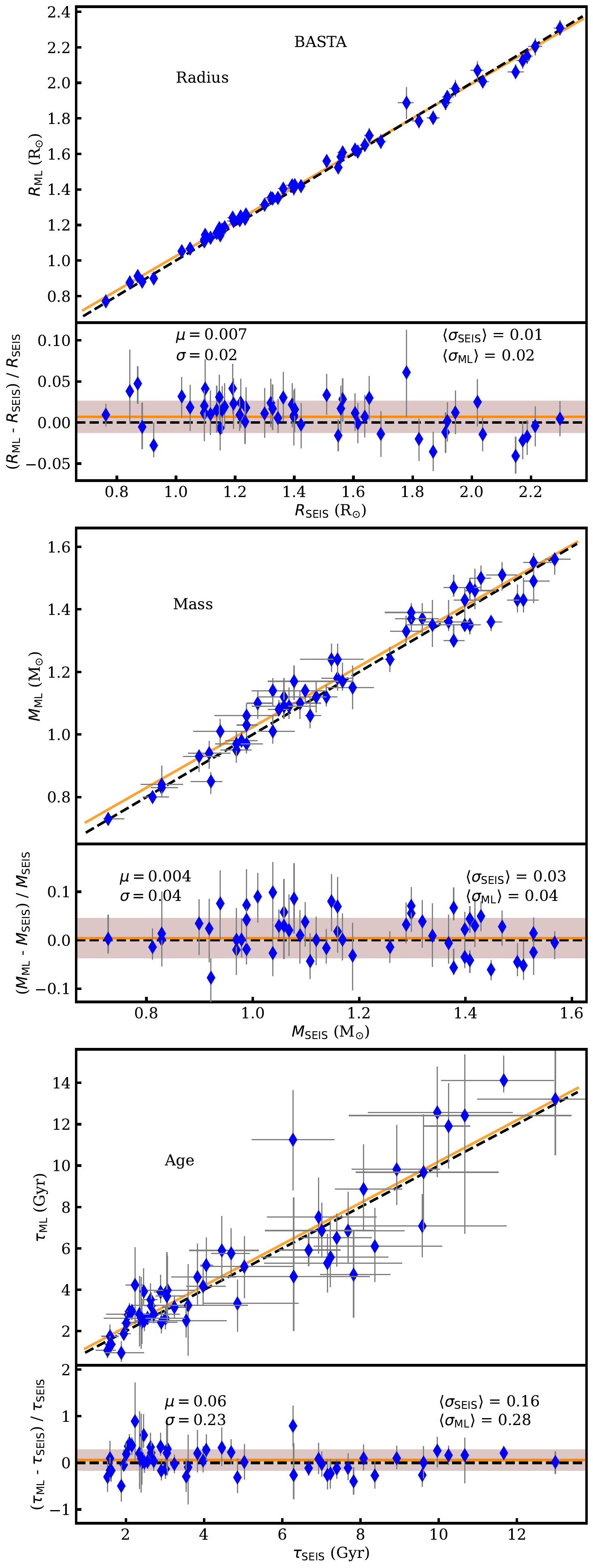}
            \caption{Same as Fig.~\ref{LEGACY_radius_mass_age}, except for results from BASTA pipeline.}
        \label{BASTA}
\end{figure}

\begin{figure}
        \centering
        \includegraphics[scale=0.33]{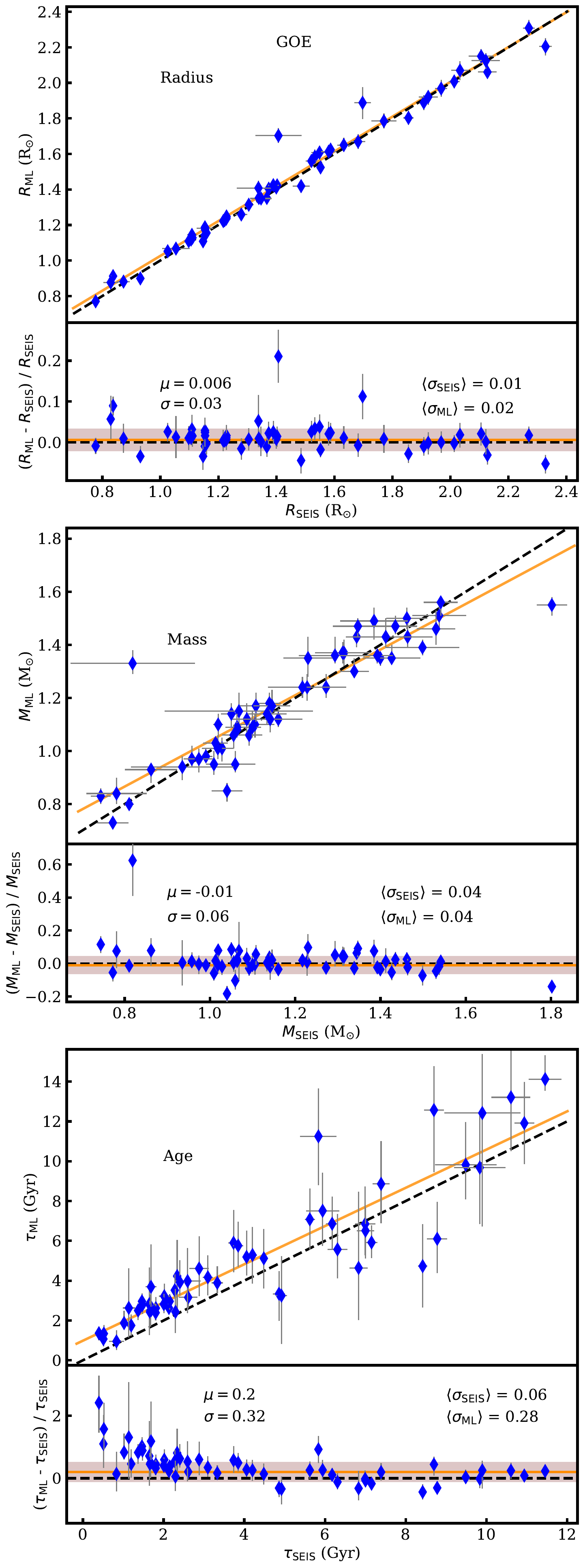}
            \caption{Same as Fig.~\ref{LEGACY_radius_mass_age}, except for results from  GOE pipeline.}
        \label{GOE}   
\end{figure}

We conducted additional comparisons between the  estimates from MAISTEP and those derived from other pipelines in the LEGACY analyses. Here, we report only the results from the two pipelines that show the best and worst agreement in age compared to MAISTEP. The BAyesian STellar Algorithm \citep[BASTA;][]{silva2015ages} and GOEttingen \citep[GOE;][]{appourchaux2015seismic} pipelines yield consistent estimates of radius and mass in comparison to MAISTEP, with  bias (and scatter) values of 0.7\% (2\%) and 0.6\% (3\%), 0.4\% (4\%) and -1\% (6\%), as illustrated in the top and middle panels of Fig.~\ref{BASTA} and Fig.~\ref{GOE}, respectively. In age, we find a slight disagreement with bias (and scatter) values of 6\% (23\%) and 20\% (32\%) as shown in the bottom panels of Fig.~\ref{BASTA} and Fig.~\ref{GOE}, respectively. We note that the primary difference in the results obtained using BASTA and GOE stem from the sets of constraints employed during optimisation: frequency ratios in BASTA and individual oscillation frequencies for GOE.

\subsection{Comparison of results from two sets of constraint}
\begin{figure}
        \centering
        \includegraphics[scale=0.33]{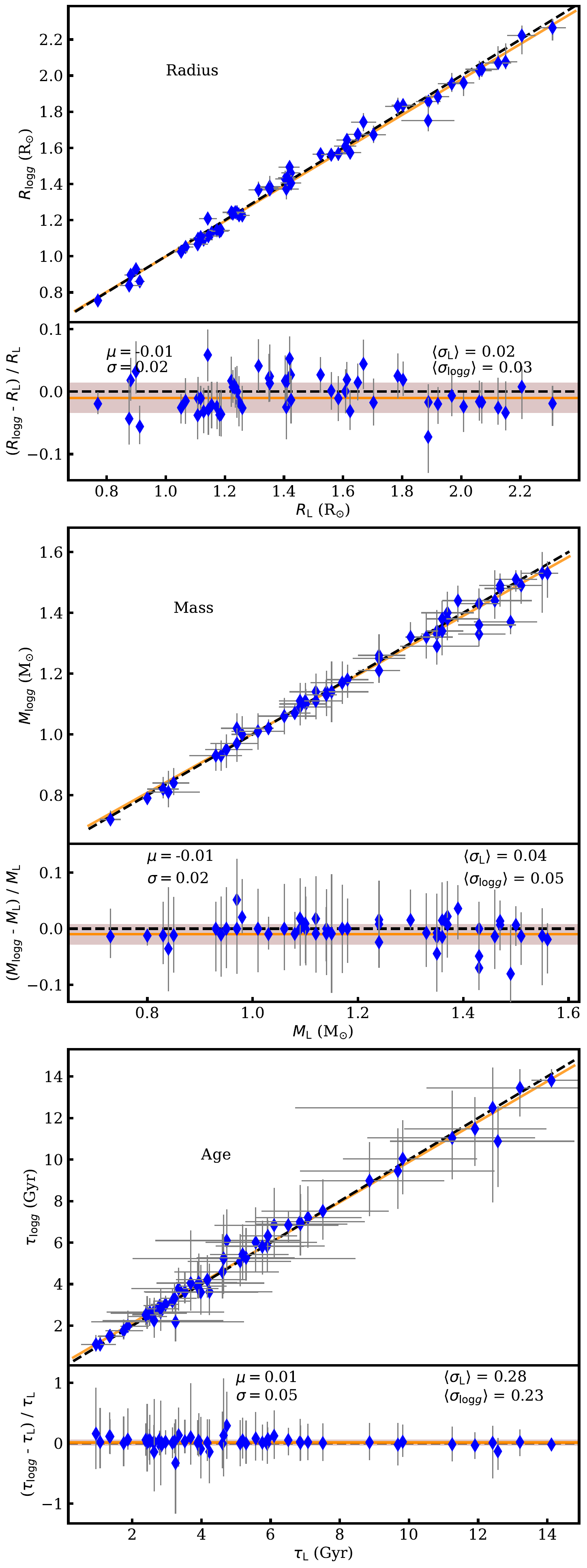}
            \caption{Same as Fig.~\ref{LEGACY_radius_mass_age}, except for the combinations of $T_{\text{eff}}$, [Fe/H], and log $g$ versus $T_{\text{eff}}$, [Fe/H], and $L$ as constraints.}
        \label{lum_logg}   
\end{figure}

We performed an additional test by training our algorithms with $T_{\text{eff}}$, [Fe/H] and log $g$, instead of $T_{\text{eff}}$, [Fe/H], and $L$. Applying the model to the LEGACY sample, we obtain consistent results as shown in Fig.~\ref{lum_logg}. The resulting bias (and scatter) is -1\% (2\%) in radius, -1\% (2\%) in mass, and 1\% (5\%) in age.  

\label{lastpage}

\end{document}